# Analytical expressions of variable specific yield for layered soils in shallow water table environments

Xue Xiao[1, 2], Xu Xu[1, 2]*, Chen Sun[3], Guanhua Huang[1, 2]

*A preprint* (May 07, 2020)

[1] Chinese-Israeli International Center for Research and Training in Agriculture, China Agricultural University, Beijing 100083, P. R. China

[2] Center for Agricultural Water Research, China Agricultural University, Beijing 100083, P. R. China

[3] Institute of Environment and Sustainable Development in Agriculture, Chinese Academy of Agricultural Sciences, Beijing 100081, P. R. China,

**Abstract**：This paper presents analytical expressions of variable specific yield for layered soils in shallow water table environments, with introducing two distinct concepts of point specific yield ($S_{yp}$) and interval average specific yield ($S_{yi}$). The $S_{yp}$ and $S_{yi}$ refer to the specific yield for the water table fluctuation approaching zero infinitely and that for an interval fluctuation of water table, respectively. On the basis of specific yield definition and van Genuchten model of soil water retention, the analytical and semi-analytical expressions were respectively proposed for $S_{yp}$ and $S_{yi}$ towards layered soils. The analytical expressions are evaluated and verified by experimental data and comparison with the previous expressions. Analyses indicate our expressions for $S_{yp}$ and $S_{yi}$ could effectively reflect the changes and nonlinear properties affected by soil hydraulic properties and soil layering under shallow water table conditions. The previously confused understanding of $S_{yp}$ and $S_{yi}$ are also distinguished. The practicality and applicability for the specific yield expressions are comprehensively analyzed for the potential applications related to the subsurface water modeling and management issues.

## 1. Introduction

Specific yield is a very important hydraulic parameter in estimating the groundwater recharge, drainage and evaporation and modeling the groundwater table dynamics for a phreatic aquifer (Gribovszki, 2018; Sophocleous, 1985; Xu et al., 2012). It plays a critical role in solving a number of relevant engineering problems and the management of water resources in various research fields (e.g. hydrogeology, hydrology and agricultural engineering) (Said et al., 2005;

---

＊ Email address for corresponding author: xushengwu@cau.edu.cn

Skaggs, 1980; Moriasi et al., 2009). The specific yield ($S_y$) (also called drainable porosity) is defined as the volume of water released from or taken into a soil column extending from the water table to the groundwater surface, per unit aquifer area of per unit change in the water table (Bear, 1972; Freeze & Cherry, 1979; Brutsaert, 2005), mathematically written as:

$$S_y = \frac{\Delta V}{A \Delta h} \quad (1)$$

where $A$ is an aquifer area and $\Delta V$ is the volume of water released or stored resulting from $\Delta h$ water table fluctuation. In this paper, we focus on studies of the ultimate specific yield as the soil profile changes from the initial equilibrium state to its new equilibrium state (Nachabe, 2002). This value is significant and widely used in subsurface water modeling (Harbaugh, et al., 2000; Said et al., 2005) and management issues (Gribovszki, 2018; Chinnasamy et al., 2018). Researchers have recognized that the specific yield could be seen as a constant parameter if only there is a linear relationship between water table fluctuations and released volumes holds, which may be valid for very coarse aquifer or under deep water table conditions; whereas, this relationship is nonlinear for the shallow water table aquifer. Due to the effects of capillary properties of soils, the specific yield is a variable parameter (Duke, 1972; Sophocleous, 1985; Nachabe, 2002).

Duke (1972) is the first among to note the nonlinear behavior of specific yield in the shallow water table conditions. He proposes an expression of variable specific yield with just a simple derivation, as follows:

$$S_y = \frac{[\theta_s - \theta(z)]dz}{dz} = \theta_s - \theta(d) \quad (2)$$

where $\theta_s$ is the saturated water content, $\theta(z)$ is the actual soil water content, $z$ is the vertical coordinate (positive upward), $d$ is the distance from water table to soil surface (positive, equal to water table depth (WTD)), and $\theta(d)$ is the average water content evaluated in a layer of the thickness $dz$, at the soil surface. With adopting the Brooks and Corey (B-C) model of soil water retention (SWR) (Brooks & Corey, 1964), Duke (1972) then introduces an analytical expression of variable specific yield, as follows:

$$S_y = (\theta_s - \theta_r)\left[1 - \left(\frac{h_a}{d}\right)^\lambda\right] \quad (3)$$

where $\theta_r$ is residual water content, $h_a$ is the soil air-entry pressure head (positive), and $\lambda$ is the soil property index. However, the equation (3) is only valid when WTD greater than $h_a$, due to the shortcomings of B-C model in describing the near saturated state. However, van Genuchten (VG) model can efficiently improve the description of SWR (van Genuchten, 1980) near saturation, as follows:

$$\theta(h) = \theta_r + \frac{(\theta_s - \theta_r)}{[1+|\alpha h|^n]^{(1-1/n)}} \tag{4}$$

where $h$ is the soil water pressure head (negative), and $\alpha$ and $n$ are the empirical shape parameters. $n$ determines the steepness of the curve, while $\alpha$ equals to the inverse of soil air-entry pressure head (i.e. $1/h_a$). Therefore, some researchers (Loheide et al., 2005; Tan et al., 2006) directly replace the B-C model with the VG model for describing $\theta(d)$ based on equation (2) by Duke (1972), and the $S_y$ can be expressed as:

$$S_y = \theta_s - [\theta_r + \frac{\theta_s - \theta_r}{[1+(\alpha d)^n]^{(1-1/n)}}] \tag{5}$$

Crosbie et al. (2005) also present a similar expression but use the mean value of initial and final WTDs instead of $d$ in equation (5). Furthermore, Nachabe (2002) demonstrates that Duke's expression (i.e. equation 2) is only valid for the conditions of "small" water table fluctuation. Later in this study, we will basically demonstrate that the equation (2) is only valid when water table fluctuation approaches zero infinitely but not just small. Hence, we tentatively define the expression based on equation (2) as *point specific yield* ($S_{yp}$). Nachabe (2002) introduces another analytical expression for variable specific yield using the method of characteristic (MOC) with the integration of B-C model, as follows:

$$S_y = (\theta_s - \theta_r) + \frac{(\theta_s - \theta_r)}{(1-\lambda)} \frac{h_a}{(d_2 - d_1)} \left( \left(\frac{h_a}{d_1}\right)^{\lambda-1} - \left(\frac{h_a}{d_2}\right)^{\lambda-1} \right) \tag{6}$$

where $d_1$ and $d_2$ are initial and final equilibrium WTDs, respectively. Compared to Duke's expression (equation 2), the equation (6) is able to calculate specific yield of single-layer soils for any water table fluctuation ranges with the given $d_1$ and $d_2$, when $d_1$ and $d_2$ are both larger than $h_a$. We define the similar expression of equation (6) as *interval average specific yield* ($S_{yi}$), as it describes the water release or storage for WTD changing from $d_1$ to $d_2$. The $S_{yi}$ value may be very useful in practical evaluation of groundwater recharge, drainage and evaporation. Note the equation (6) has similar flaws near saturation when WTD is less than $h_a$. Later, Cheng et al. (2015) derive a semi-analytical expression of $S_{yi}$ for single-layer soils through integrating the VG model by Taylor series expansion, with introducing the error term in different integration intervals at a given expansion point. Moreover, it is found that the concepts of $S_{yp}$ and $S_{yi}$ are easily confused or misused in some previous studies (Lei et al., 1984; Crosbie et al., 2005).

To our knowledge, the accurate expression of variable $S_y$ for layered soils is rarely reported. Meanwhile, it is very critical to clarify the concepts of $S_{yp}$ and $S_{yi}$ for avoiding the confusion use. Therefore, the objectives of this paper were: (1) to introduce the analytical expressions for point specific yield ($S_{yp}$) and interval average specific yield ($S_{yi}$) towards layered soils, on the basis of the closed-form VG model; and (2) to propose and clarify two significant concepts of $S_{yp}$ and $S_{yi}$ for accurate use of specific yield. Then, the new expressions were verified and evaluated with the theoretical comparison and experimental data. The distinction of $S_{yp}$ and $S_{yi}$ was systematically analyzed in comparison with the existing expressions. The applicability and practicality were finally presented for our proposed $S_{yp}$ and $S_{yi}$ expressions.

## 2. Theoretical statement and derivation

Taking a drainage process as an example, when the water table changes from the initial depth $d$ to the final depth $(d+\Delta d)$ with both an equilibrium status of soil water distribution above the water table, the darkened area can represent the water volume released from the soil column of unit aquifer area (Figure 1a). Thus, the total amount of water ($\Delta V$) drained out for unit aquifer area, from the initial equilibrium state (curve $S_1$) to the final equilibrium state (curve $S_2$) of soil water distribution, is calculated as (Bear, 1972; Bear & Cheng, 2010):

$$\Delta V = \int_0^{\Delta d} \theta_s dz + \int_{\Delta d}^{d+\Delta d} \theta_1(z)dz - \int_0^{d+\Delta d} \theta_2(z)dz \tag{7}$$

where $\theta_1(z)$ and $\theta_2(z)$ are the functions for soil water content in initial and final equilibrium states, respectively. According to the equations (1) and (7), the general expression of specific yield is:

$$S_y = \frac{\Delta V}{A\Delta d} = \frac{\int_0^{\Delta d} \theta_s dz + \int_{\Delta d}^{d+\Delta d} \theta_1(z)dz - \int_0^{d+\Delta d} \theta_2(z)dz}{\Delta d}$$
$$= \theta_s + \frac{\int_{\Delta d}^{d+\Delta d} \theta_1(z)dz - \int_0^{d+\Delta d} \theta_2(z)dz}{\Delta d} \tag{8}$$

The mathematical derivation of analytic expressions of specific yield (i.e. for $S_{yp}$ and $S_{yi}$) is given in detail in the following parts.

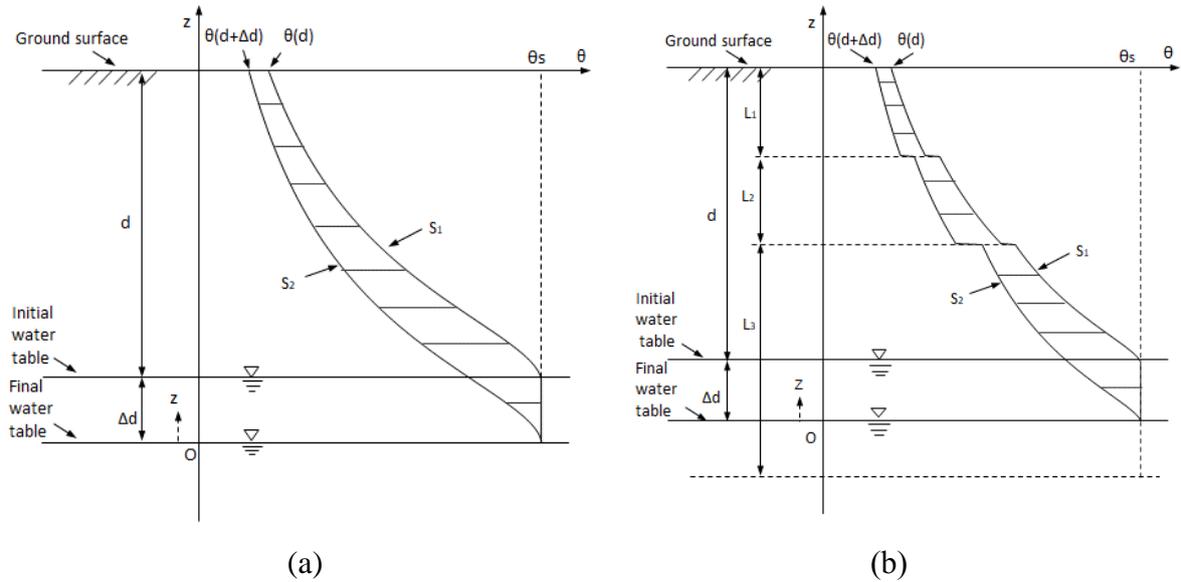

(a)                    (b)

**Figure 1.** Water table changes from the initial depth $d$ to the final depth $(d+\Delta d)$ with both an equilibrium status of soil water distribution above the water table for: single-layer (homogeneous) soil column (a), and layered soil column with three layers as an example (b).

*2.1. Point specific yield*

When assuming $\Delta d$ is small enough and approaches to zero infinitely, the equation (8) can be rewritten as:

$$S_{yp} = \lim_{\Delta d \to 0} \frac{\Delta V}{A \Delta d} = \lim_{\Delta d \to 0} \frac{\int_0^{\Delta d} \theta_s dz + \int_{\Delta d}^{d+\Delta d} \theta_1(z)dz - \int_0^{d+\Delta d} \theta_2(z)dz}{\Delta d}$$
$$= \theta_s + \lim_{\Delta d \to 0} \frac{\int_{\Delta d}^{d+\Delta d} \theta_1(z)dz - \int_0^{d+\Delta d} \theta_2(z)dz}{\Delta d} \quad (9)$$

We define the specific yield expressed in equation (9) as the *point specific yield* ($S_{yp}$), due to that the expression and the related subsequent derivation are based on the limit concept (i.e. $\Delta d \to 0$). $g_1(z)$ and $g_2(z)$ are defined as the primitive functions for $\theta_1(z)$ and $\theta_2(z)$, respectively, i.e., $g'_1(z) = \theta_1(z)$ and $g'_2(z) = \theta_2(z)$. The functions of $g_1(z)$ and $\theta_1(z)$ should satisfy the below mathematical relation:

$$\theta_1(z) = \lim_{\Delta z \to 0} \frac{g_1(z + \Delta z) - g_1(z)}{\Delta z} \quad (10)$$

This is similar for $g_2(z)$ and $\theta_2(z)$ as well. In addition, the functions $\theta_1(z)$ and $\theta_2(z)$ satisfy the relation of $\theta_1(z-\Delta d) = \theta_2(z)$, resulting in $g_1(z-\Delta d) = g_2(z)$. Referring to Lei et al. (1984), the equation (9) can be rewritten as:

$$\begin{aligned} S_{yp} &= \theta_s + \lim_{\Delta d \to 0} \frac{[g_1(d+\Delta d) - g_1(\Delta d) - g_2(d+\Delta d) + g_2(0)]}{\Delta d} \\ &= \theta_s + \lim_{\Delta d \to 0} \frac{[g_2(d) - g_2(0) - g_2(d+\Delta d) + g_2(0)]}{\Delta d} \\ &= \theta_s - \lim_{\Delta d \to 0} \frac{[g_2(d+\Delta d) - g_2(d)] + [g_2(0) - g_2(0)]}{\Delta d} \\ &= \theta_s - \theta_2(d) \end{aligned} \quad (11)$$

which is exactly equation (2) provided by Duke (1972). It is worth noting that the equation (11) is obtained based on a rigorous derivation with an assumption of $\Delta d \to 0$. This implies that it only can be used as an approximation of specific yield for a very small water fluctuation.

Replacing $\theta_2(d)$ in the last term of right side in equation (11) by equation (4), the point specific yield for single-layer (homogeneous) soils (noted as $S_{yp,sl}$) can be finally expressed as:

$$S_{yp,sl} = (\theta_s - \theta_r)\left\{1 - [1 + (\alpha d)^n]^{(1/n-1)}\right\} \quad (12)$$

The equation (12) is an expression of variable specific yield similar to equation (5), with the rigorous mathematical derivation and an assumption of $\Delta d \to 0$. For layered soils, the $S_{yp}$ value is further affected by the vertical stratification of soils above the water table. Figure 1b presents the equilibrium state of soil water profile (curve $S_1$) in initial depth $d$ and that (curve $S_2$) in final depth ($d+\Delta d$), with taking 3 soil layers as an example. The darkened area represents the released water volume. According to the equation (1), the point specific yield for a 3-layer soil column can be expressed as:

$$S_{yp,ml} = \lim_{\Delta d \to 0} \frac{\Delta V}{A \Delta d}$$

$$= \lim_{\Delta d \to 0} \frac{(\int_{d+\Delta d-L_1}^{d+\Delta d} \theta_{11} dz - \int_{d+\Delta d-L_1}^{d+\Delta d} \theta_{12} dz) + (\int_{d+\Delta d-L_1-L_2}^{d+\Delta d-L_1} \theta_{21} dz - \int_{d+\Delta d-L_1-L_2}^{d+\Delta d-L_1} \theta_{22} dz)}{\Delta d}$$

$$+ \lim_{\Delta d \to 0} \frac{+(\int_0^{\Delta d} \theta_{3,s} dz + \int_{\Delta d}^{d+\Delta d-L_1-L_2} \theta_{31} dz - \int_0^{d+\Delta d-L_1-L_2} \theta_{32} dz)}{\Delta d}$$
(13)

where $S_{yp,ml}$ is the point specific yield of layered soils, and $\theta_{11}$, $\theta_{21}$ and $\theta_{31}$ are the functions for soil water content in initial equilibrium state (with WTD of $d$) for the 1st, 2nd and 3rd layer, respectively. $\theta_{12}$, $\theta_{22}$ and $\theta_{32}$ are the functions for soil water content in final equilibrium state (WTD of $d+\Delta d$) for the 1st, 2nd and 3rd layer, respectively. $\theta_{3,s}$ is the saturated soil water content for the 3rd layer. The initial WTD $d \in [L_1+L_2, L_1+L_2+L_3]$.

Similar to the derivation in equation (11), the equation (13) can be rewritten as:

$$S_{yp,ml} = \theta_1(d-L_1) - \theta_1(d) + \theta_2(d-L_1-L_2) - \theta_2(d-L_1) + \theta_{3,s} - \theta_3(d-L_1-L_2) \quad (14)$$

Generalizing this expression to the $k$-layer soils, one can be:

$$\begin{cases} S_{yp,ml} = \sum_{j=1}^{k-1}[\theta_j(d-\sum_{i=1}^{j}L_i) - \theta_j(d-\sum_{i=1}^{j-1}L_i)] + \theta_{k,s} - \theta_k(d-\sum_{i=1}^{k-1}L_i) \\ \theta_j(x) = \theta_{r,j} + (\theta_{s,j} - \theta_{r,j})/[1+|\alpha_j x|^{n_j}]^{(1-1/n_j)} \end{cases} \quad (15)$$

where $\theta_j$ is the SWR function for the $j$-th layer (calculated by equation 4 with VG model), $L_i$ is the thickness for the $i$-th layer, $k$ is number of soil layers above the water table, and $x$ refers to an independent variable. Therefore, the equation (15) is a new, generalized analytical expression of point specific yield for layered soils.

*2.2. Interval average specific yield*

The equation (12) or (15) is limited to approximately calculate specific yield for very small WTD changes. Based on the equations (1) and (12), the interval average specific yield for single-layer soil (Figure 2a) with large water table fluctuations can be expressed as:

$$S_{yi,sl} = \frac{\Delta V}{A \Delta d} = \frac{\int_{d_1}^{d_2} S_{yp} dz}{\int_{d_1}^{d_2} dz} = \frac{\int_{d_1}^{d_2}(\theta_s - \theta_r)\{1-[1+(\alpha z)^n]^{(1/n-1)}\}dz}{d_2 - d_1}$$

$$= (\theta_s - \theta_r) - \frac{(\theta_s - \theta_r)\int_{d_1}^{d_2}[1+(\alpha z)^n]^{(1/n-1)}dz}{d_2 - d_1}$$
(16)

The formula in the integral sign of equation (16) is highly nonlinear, and thus it is difficult to obtain the integral expression directly. In this study, we derive an approximate analytical expression for $S_{yi}$ using the compound Simpson's rule (Davis & Rabinowitz, 1984) in the following part.

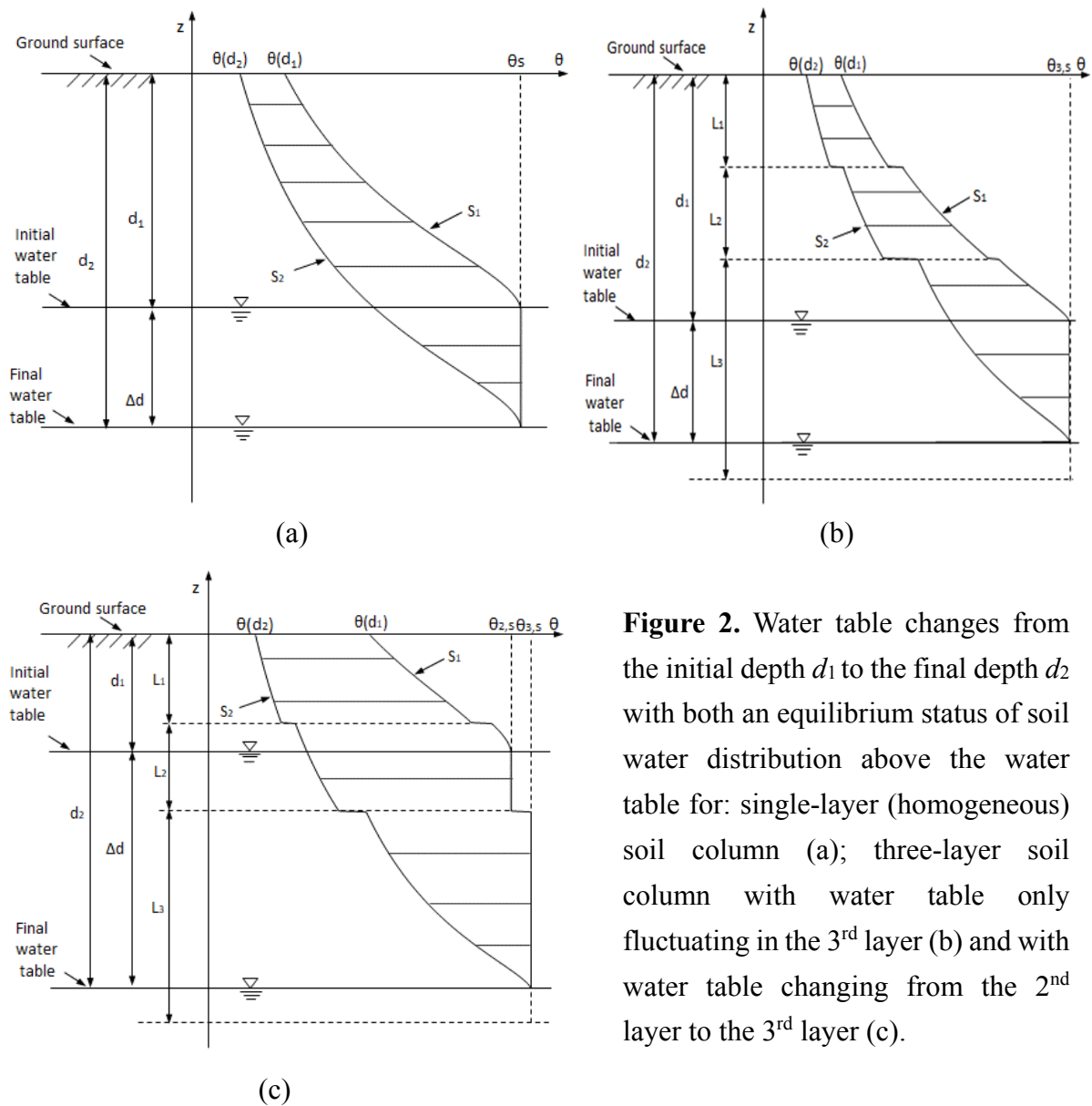

**Figure 2.** Water table changes from the initial depth $d_1$ to the final depth $d_2$ with both an equilibrium status of soil water distribution above the water table for: single-layer (homogeneous) soil column (a); three-layer soil column with water table only fluctuating in the 3$^{rd}$ layer (b) and with water table changing from the 2$^{nd}$ layer to the 3$^{rd}$ layer (c).

Here we set $f(z) = [1+(\alpha z)^n]^{(1/n-1)}$. Let

$$d_1 = z_0 < z_1 < \cdots < z_m = d_2$$

be a sequence of equal subintervals which divide the interval $[d_1, d_2]$ into $m$ parts, with $l = \frac{d_2 - d_1}{m}$ and marking $z_{i+1/2} = z_i + \frac{1}{2}l$, $i = 0,1,2,\cdots,m-1$. Then by the compound Simpson's rule, we can obtain:

$$I = \int_{d_1}^{d_2} [1+(\alpha z)^n]^{(1/n-1)} dz = \sum_{i=0}^{m-1} \int_{z_i}^{z_{i+1}} f(z)dz \qquad (17)$$
$$= \frac{l}{6}[f(d_1) + 4\sum_{i=0}^{m-1} f(z_{i+1/2}) + 2\sum_{i=1}^{m-1} f(z_i) + f(d_2)] - \frac{(d_2-d_1)}{180}(\frac{l}{2})^4 f^{(4)}(\eta), \quad \eta \in (d_1, d_2)$$

Meanwhile, the compound Simpson's rule approximation term is:

$$S_m = \frac{l}{6}[f(d_1) + 4\sum_{k=0}^{m-1} f(z_{i+1/2}) + 2\sum_{k=1}^{m-1} f(z_i) + f(d_2)] \qquad (18)$$

The remainder term (error term) is:

$$R_m = -\frac{(d_2-d_1)}{180}(\frac{l}{2})^4 f^{(4)}(\eta), \quad \eta \in (d_1, d_2) \qquad (19)$$

where $m$ is the number of equal subintervals divided by interval $[d_1, d_2]$, $l$ is the length of each subinterval, $S_m$ is the compound Simpson's rule approximation term, and $R_m$ is the remainder term.

Substituting equations (18) and (19) into equation (16), then we obtain:

$$S_{yi,sl} = (\theta_s - \theta_r) - \frac{(\theta_s - \theta_r)}{(d_2 - d_1)}\left\{\frac{l}{6}[f(d_1) + 4\sum_{k=0}^{m-1} f(z_{i+1/2}) + 2\sum_{k=1}^{m-1} f(z_i) + f(d_2)] - \frac{(d_2-d_1)}{180}(\frac{l}{2})^4 f^{(4)}(\eta)\right\} \qquad (20)$$
$$, \eta \in (d_1, d_2)$$

This is the new analytical expression of the variable $S_{yi}$ for single-layer soils. If ignoring the error term, the approximation expression is:

$$S_{yi,sl} = (\theta_s - \theta_r) - \frac{(\theta_s - \theta_r)}{(d_2 - d_1)}\left\{\frac{l}{6}[f(d_1) + 4\sum_{k=0}^{m-1} f(z_{i+1/2}) + 2\sum_{k=1}^{m-1} f(z_i) + f(d_2)]\right\} \qquad (21)$$

Substituting function $f(z)$ expression into equation (21), the approximate analytical expression for the specific yield in interval $[d_1, d_2]$ can be expressed as:

$$S_{yi,sl} = (\theta_s - \theta_r) - \frac{(\theta_s - \theta_r)}{6m}\left\{\begin{array}{l}[1+(\alpha d_1)^n]^{(1/n-1)} + 4\sum_{k=0}^{m-1}[1+(\alpha z_{i+1/2})^n]^{(1/n-1)} + 2\sum_{k=1}^{m-1}[1+(\alpha z_i)^n]^{(1/n-1)} \\ +[1+(\alpha d_2)^n]^{(1/n-1)}\end{array}\right\} \qquad (22)$$

The total remainder term ($TR_m$) (i.e. the error term) is:

$$TR_m = \frac{(\theta_s - \theta_r)(d_2 - d_1)^4}{2880 m^4} f^{(4)}(\eta), \qquad \eta \in (d_1, d_2) \tag{23}$$

The number of subintervals $m$ depends on the accuracy required. Thus, the equation (22) is a new semi-analytical expression of $S_{yi}$ for single-layer soils.

When calculating the $S_{yi}$ for layered soils, there are generally two typical situations: one is for the water table only fluctuating in one of the layers (e.g. only in the 3$^{rd}$ layer in Figure 2b); and the other is for water table fluctuating across different soil layers (e.g. dropping from the 2$^{nd}$ layer to the 3$^{rd}$ layer, as shown in Figure 2c). For the case of the former situation (Figure 2b), the interval average specific yield for layered soils ($S_{yi,ml}$) can be expressed as:

$$S_{yi,ml} = \frac{\Delta V}{A \Delta d} = \frac{\int_{d_1}^{d_2} S_{yp,ml,3} dz}{\int_{d_1}^{d_2} dz} \tag{24}$$

The $S_{yp,ml,3}$ refers to the $S_{yp,ml}$ for the 3-layer soils, which is expressed by equation (14). For the case of the latter situation (Figure 2c), the $S_{yi,ml}$ can be expressed as:

$$S_{yi,ml} = \frac{\Delta V}{A \Delta d} = \frac{\sum_{i=2}^{3} \Delta V_i}{\int_{d_1}^{d_2} dz} = \frac{\int_{d_1}^{L_1+L_2} S_{yp,ml,2} dz + \int_{L_1+L_2}^{d_2} S_{yp,ml,3} dz}{d_2 - d_1} \tag{25}$$

More generally, the $S_{yi,ml}$ for $k$-layer soils can be described, as follows:

$$S_{yi,ml} = \frac{\Delta V}{A \Delta d} = \frac{\sum_{i=2}^{k} \Delta V_i}{\int_{d_1}^{d_2} dz} = \frac{\int_{d_1}^{L_1+L_2} S_{yp,ml,2} dz + \int_{L_1+L_2}^{L_1+L_2+L_3} S_{yp,ml,3} dz + \cdots}{d_2 - d_1}$$

$$+ \frac{\int_{L_1+L_2+\cdots+L_{n-2}}^{L_1+L_2+\cdots+L_{n-1}} S_{yp,ml,n-1} dz + \int_{L_1+L_2+\cdots+L_{n-1}}^{d_2} S_{yp,ml,n} dz}{d_2 - d_1} \tag{26}$$

where the $S_{yp,ml,i}$ refers to the $S_{yp,ml}$ for the $i$-layer soils, which is expressed by equation (15). Therefore, the $S_{yi,ml}$ can be easily calculated by equation (26), with the same procedures applied for $S_{yi,sl}$ calculation (equations 17 to 23).

## 3. Analysis and discussion

### 3.1. Point specific yield: theoretical comparison and applicability

A calculation case was conducted to show the performance of the point specific yield expression for layered soils (equation 15) under various soil conditions, aiming to reveal and analyze the effects of soil hydraulic properties and soil layering on the $S_{yp,ml}$. In the case, we assumed six soil columns with combination of different soil textures and layering conditions (soil columns 1 to 6 in Figure 3). Three typical soil types of silt loam, loam, and sandy loam, reported by Carsel and Parrish (1988) were considered in this case. The soil hydraulic parameters were provided in detail in Table 1.

The variations of $S_{yp}$ were calculated with WTD increasing from top surface to bottom for these soil columns (Figure 4). Results showed that the $S_{yp}$ varied nonlinearly with the increase of WTD, with greater values for coarser soils. It was obvious that the coarser soil column had a sharper increasing trend, showing stronger nonlinear behaviors (soil columns 1 to 3 in Figure 4). For the layered soils, the $S_{yp}$ curves had a sudden change when the soil texture changed between layers (columns 4 to 6 in Figure 4); moreover, with the increase of WTD, the $S_{yp}$ values were gradually close to and even larger than those for the single-layer soil (column 3 with loamy sand). In addition, the thickness of the soil layer should be a significant impact factor for the $S_{yp}$. For example, in our case, changing the thickness (i.e. 30, 55 and 80 cm) of the middle layer (loam) resulted in marked differences among the three $S_{yp,ml}$ curves (columns 4 to 6, Figure 4). The thinner the middle layer was, the faster the above curve approached to the column 3 curve. The thickest layer (i.e. the bottom layer in this case) would play a primary role in $S_{yp,ml}$ variation. Overall, the $S_{yp}$ was closely related to the soil hydraulic properties and soil layering above the water table. Furthermore, above results have indicated that the $S_{yp}$ varied much with the WTD and the soil properties for layered soils (Figures 4), and thus including the variable specific yield was necessary for improving the modeling accuracy (e.g. for groundwater flow models, hydrological models and farmland drainage models). Due to that the $S_{yp}$ can be calculated based on a given WTD, the $S_{yp}$ expression can be adopted in simulation models to estimate the specific yield for the next time-step and to drive the model for layered soils. In addition, the $S_{yp,ml}$ curve was continuous and smooth, and it may not affect the convergence of the model in iterative calculation.

Moreover, the $S_{yp}$ expression was rigorously derived based on the limit concept of $\Delta d \to 0$, and it was only the approximation of specific yield for very small groundwater fluctuations. Thus, we recommended a smaller calculation time-step when using the $S_{yp}$ expressions in modeling. Particularly, due to a strong nonlinearity for $S_{yp}$ in near saturated state of soil profile, the time-step should be very small when water table was too shallow.

*3.2. Interval average specific yield: verification and practicality*

*3.2.1. Verification of the expressions*

(1) Experimental verification

 *(a) Case for single-layer soil*

The drainage experimental data with a homogenous sand soil column (available from Cheng et al. 2015) was adopted to test the proposed expression of $S_{yi,sl}$ (equation 22). The soil hydraulic parameters for the B-C model and VG model were listed in Table 1. In the experiment, the drainage events were conducted six times discontinuously, with WTD changing from 0 to

18.7 cm, 20.3 to 33.1 cm, 33.9 to 51.3 cm, 51.9 to 64 cm, 64.5 to 79.5 cm, and 80.2 to 90 cm. The measured $S_{yi,sl}$ values were obtained according to the drained water volume and the measured WTD fluctuations. The detailed description of the experiment can be found in Cheng et al. (2015). The calculated values were determined with equation (22) and equation (6) (i.e. Nachabe's expression), respectively.

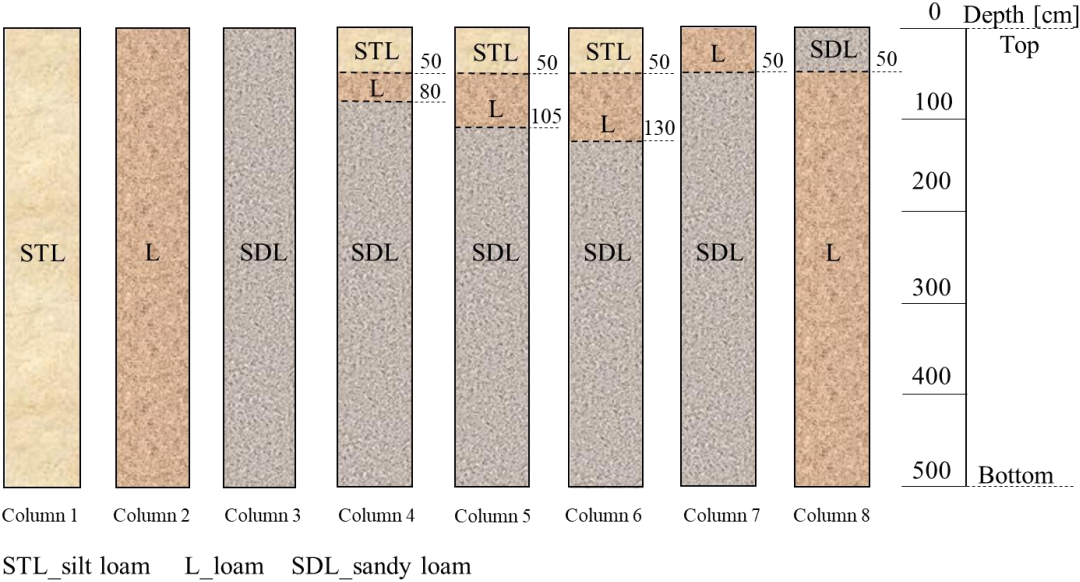

**Figure 3.** Schematic diagram of different soil columns with single, two or three soil layers in case scenarios. The soil hydraulic parameters were adopted from Carsel and Parrish (1988) (Table 1).

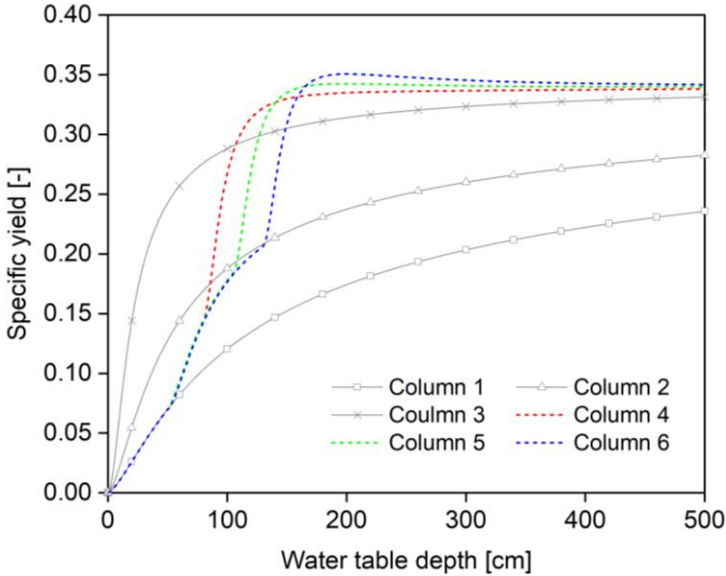

**Figure 4.** The variations of point specific yield ($S_{yp}$) for the single-layer and layered soil columns (using 3 layers as an example), with soil hydraulic parameters available from Carsel and Parrish (1988).

Table 1 Soil types and the parameters of soil water retention curves adopted in soil column cases.

| Soil types | Brooks-Corey model | | | | van Genuchten model | | | |
|---|---|---|---|---|---|---|---|---|
| | $\theta_s$ (cm³ cm⁻³) | $\theta_r$ (cm³ cm⁻³) | $\lambda$ (-) | $h_a$ (cm) | $\theta_s$ (cm³ cm⁻³) | $\theta_r$ (cm³ cm⁻³) | $\alpha$ (cm⁻¹) | $n$ (-) |
| Sand[1] | 0.364 | 0.095 | 2.4 | 33.5 | 0.364 | 0.095 | 0.020 | 5.531 |
| Silt loam[2] | | | | | 0.450 | 0.067 | 0.020 | 1.41 |
| Loam[2] | | | --- | | 0.430 | 0.078 | 0.036 | 1.56 |
| Sandy loam[2] | | | | | 0.410 | 0.065 | 0.075 | 1.89 |
| Sand[3] | | | | | 0.390 | 0.30 | 0.096 | 2.60 |
| Sandy loam[3] | | | --- | | 0.420 | 0.090 | 0.013 | 1.80 |

Note: The superscripts 1 and 2 represent the soil data taken from Cheng et al. (2015) and Carsel and Parrish (1988), respectively.
The superscript 3 means that the soil parameters are obtained through fitting the measured soil water retention curves in Lei et al. (1984).

The comparison of the calculated and measured $S_{yi,sl}$ are presented in Figure 5(a). Results indicated that the equation (22) had a higher accuracy than equation (6) in this case. The calculated values by equation (22) (with RMSE=0.013) were obviously closer to the measured ones, compared with those by equation (6) (with RMSE=0.033). Moreover, when WTD less than the air-entry pressure ($d < h_a$), the proposed equation was able to calculate the $S_{yi}$, but the equation (6) lacked the description of $S_{yi,sl}$. Therefore, the equation (22) may significantly improve accuracy of $S_{yi,sl}$ and complement the description in the near saturation segment ($d < h_a$), compared to the previous Nachabe's expression.

*(b) Case for layered soils*

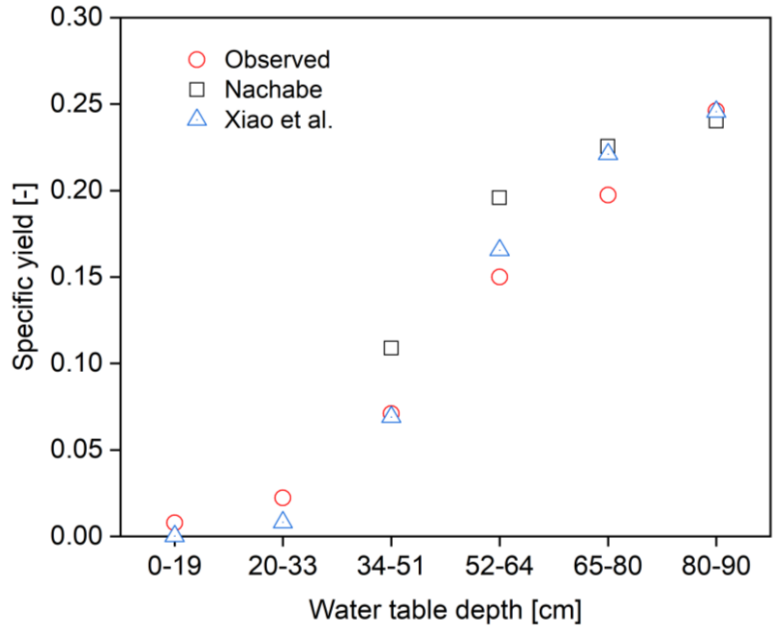

(a)

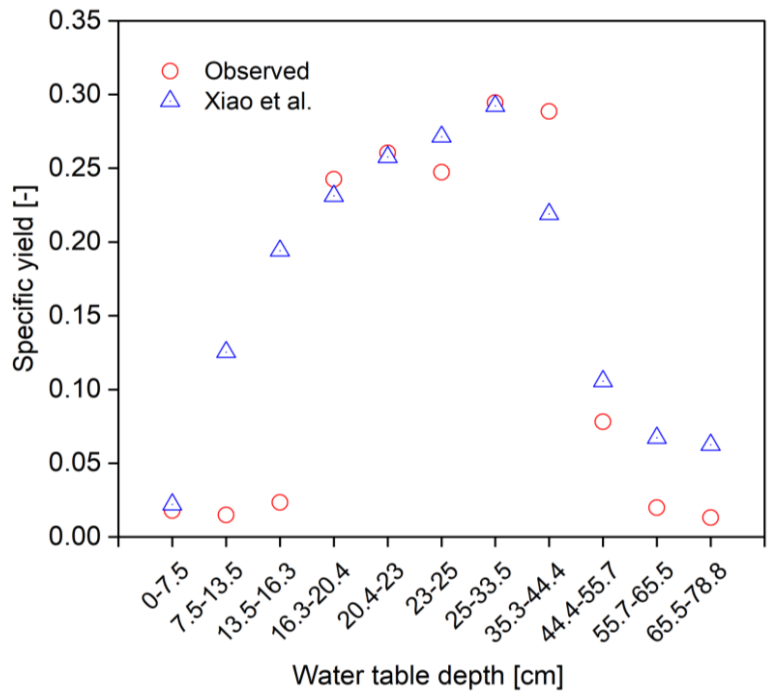

(b)

**Figure 5.** Experimental verification of the new $S_{yi}$ expressions for the single-layer soil (a) and the layered soil (b).

The drainage experimental data with a two-layer soil column (available from Lei et al., 1984) was adopted to test the proposed $S_{yi,ml}$ expression (equation 26). The soil column consisted of the sandy soil in the upper layer (0-30.6 cm) and the sandy loam soil in the lower layer (30.6-85 cm), with a diameter of 10.4 cm. The soil hydraulic parameters for VG model were obtained by fitting the soil water retention curves (Table 1). The drainage water was measured with WTD increasing from top surface to bottom (about 5-10 cm each time), and the measured $S_{yp,ml}$ values were calculated with the drained water volume and the measured WTD fluctuations.

The comparison of the calculated and measured $S_{yi,ml}$ are presented in Figure 5(b), producing a RMSE of 0.069. Results showed that the overall change trend was in agreement with the measured values, which reflecting the obvious non-linearity of $S_y$ changes for layered soils. The calculated values matched well with the measured values when WTDs changing in the middle parts of the soil column (e.g. at the depth interval of 16.3-55.7 cm). The errors between the measured and calculated were relatively large when WTDs near the surface or the bottom. This may be caused by the effect of column boundary or scale effect (the diameter was only about 10 cm) which affected the accuracy of measured values. Above comparisons indicated that the proposed $S_{yi,ml}$ expressions can be extended to well describe $S_{yi}$ changes for layered soils.

(2) Theoretical verification

Theoretically, the $S_{yi}$ should be approximately equal to the $S_{yp}$ if the water table fluctuation was small enough. Based on this, the correctness of semi-analytical expressions of $S_{yi}$ (equations 22 and 26) can be verified by the analytical expressions of $S_{yp}$ (equations 12 and 15). Here, we assumed two soil columns with the combination of two soil types (i.e. loam and sandy loam) and different layering conditions (soil columns 7 and 8, Figure 3). The variations of $S_{yp}$ with WTD increasing from top surface to bottom were calculated for the two columns. The $S_{yi,ml}$ was calculated with assuming a 0.1 cm water table fluctuation compared to initial WTD. Results showed that the calculated $S_{yi,ml}$ values almost exactly matched with the $S_{yp,ml}$ curves for both columns 7 and 8 (Figure 6). The above comparison further validated the correctness of the proposed $S_{yi,ml}$ expression.

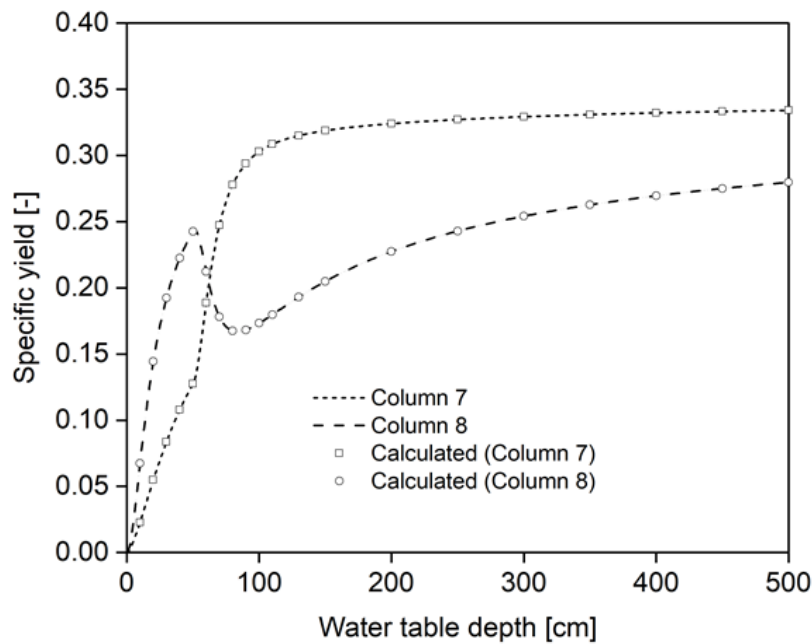

**Figure 6.** Verification of the accuracy of semi-analytical $S_{yi,ml}$ expression (equation 26, circle) by comparing with the analytical $S_{yp,ml}$ expression (equation 15, dash line) in two different soil columns.

*3.2.2. A family of $S_{yi}$ curves*

A family of $S_{yi}$ curves with water table fluctuations was proposed to show the practicality and differences of our new $S_{yi}$ expressions in recharge/drainage estimation under shallow water table conditions. Here we took the single-layer soil (i.e. soil columns 2 and 3) and layered soil (i.e. soil column 7) as examples. Considering the water table fluctuations with different initial WTDs, the $S_{yi}$ can be calculated with different final WTDs, and then a family of $S_{yi}$ curves can be drawn on the basis of equations 22 and 26.

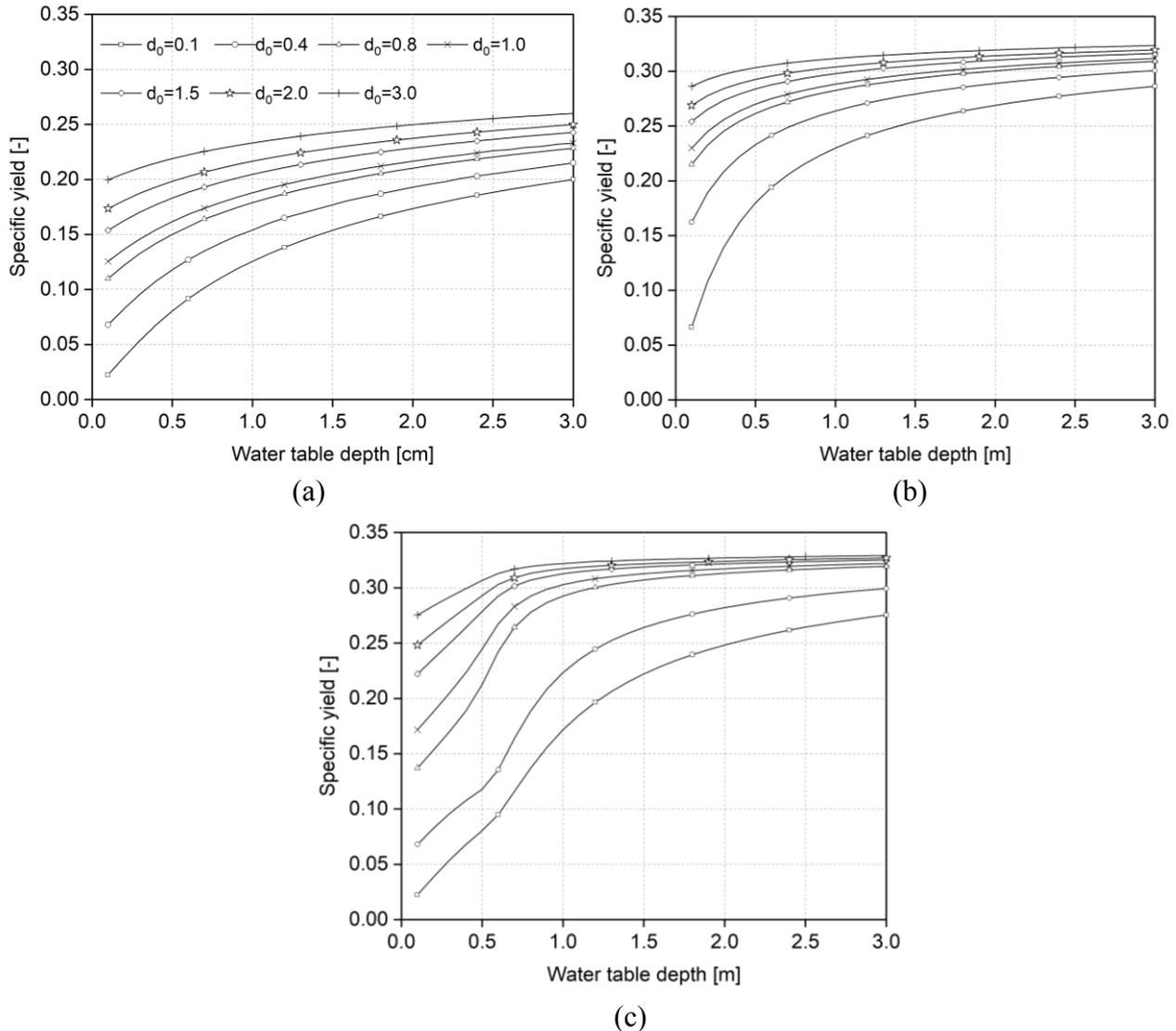

**Figure 7.** A family of $S_{yi}$ curves with water table fluctuations: (a) and (b) for single-layer soil of column 2 and 3, respectively; and (c) for layered soil of column 7.

Figure 7 showed the family of $S_{yi}$ curves for both the single-layer soil and the layered soil. It was able to query the $S_{yi}$ value with the family curves. For example, if the WTD varied from the initial $d$ = 0.4 m to the final $d$ =1.5 m (Figure 7a), one could easily obtained $S_{yi}$ = 0.177 for this fluctuation event when looking along the single line of initial WTD $d_0$ = 0.4 m. The recharge/drainage water volume can be estimated according to the initial and final WTDs and the queried $S_{yi}$ value. To inquire a family of $S_{yi}$ curves may be a very practical way for recharge/drainage estimation in shallow water table environments, especially for the highly nonlinear $S_{yi}$ under layered soil conditions (Figure 7c).

*3.3. Discussion*

It was very necessary to distinguish the point specific yield and interval average specific yield, in order to avoid confusion between them. The $S_{yp}$ expression was rigorously derived based on $\Delta d \rightarrow 0$, which was acceptable for a very small WTD fluctuation; whereas, the $S_{yi}$ expression was suitable for any water table fluctuations. While, it was possible to be confusedly understood in some previous studies. For example, the $S_{yp}$ was directly used to compare with the measured specific yield values (often with significant WTD fluctuations in the experiments); actually, these measured values should correspond to the $S_{yi}$ values. In addition, the $d$ in equation 12 could not be simply substituted by $(d_1+d_2)/2$ for calculating the specific yield for WTD fluctuations from $d_1$ to $d_2$. In fact, there should be a numerical error term due to non-linearity of specific yield. The $S_{yp}$ should be a continuous curve calculated by giving a WTD (e.g. in Figure 4), which had potential use in subsurface water modeling. The $S_{yi}$ was a series of discrete points calculated by the given initial and final WTDs (e.g. in Figure 5), which had a practical usability in estimating groundwater recharge, drainage, evaporation, etc. In addition, the use of the proposed expressions of $S_{yp}$ and $S_{yi}$ was limited in some conditions where it was hard to reach the equilibrium state of soil water profile above the water table, e.g., under very strong evaporation climates or very low permeability soils.

**4. Conclusion**

This article introduced the variable specific yield expressions for layered soil under shallow water table environments. In order to distinguish the previously confused understanding of specific yield, two significantly different concepts were defined: the point specific yield ($S_{yp}$) and the interval average specific yield ($S_{yi}$). The analytical expression of $S_{yp}$ was derived by a limited assumption of $\Delta d \rightarrow 0$, presenting a smooth and continuous curve for $S_{yp}$. The scenario comparison of six different soil columns (single-layer and layered) were presented, and the $S_{yp}$ for layered soils showed a highly nonlinear behaviors with the increase of WTD. The $S_{yp,ml}$ curve had a sudden change when the soil texture changed between layers. The thickness of the soil layer should be a significant impact factor for the $S_{yp,ml}$. Overall, the $S_{yp}$ was closely related to the WTD, soil hydraulic properties, and soil layering above the water table. The $S_{yp}$ expressions could be very useful to describe variable specific yield of layered soils for subsurface water modeling.

The compound Simpson's rule was adopted to derive the semi-analytical expression of $S_{yi}$ for layered soils. The new $S_{yi}$ expressions referred to an interval fluctuation of water table. Comparisons with experimental data and $S_{yp,ml}$ curve indicated that the proposed $S_{yi,ml}$ expressions can be extended to well describe $S_{yi}$ changes for layered soils. The new expression also improved the accuracy of Nachabe's expression. Moreover, a family of $S_{yi}$ curves were presented for both the homogenous soil and the layered soil, respectively. Results suggested that the $S_{yi}$ expressions could be necessary and useful in groundwater recharge/drainage estimation in shallow water table environments. In addition, the distinction of $S_{yp}$ and $S_{yi}$ and some misunderstanding of them were clarified for avoiding misuse in the future research.

**Acknowledgements**

This research was supported by the National Science Foundation of China (grant numbers: 51679235 and 51639009).